\documentclass[aps,
twocolumn,
showpacs,preprintnumbers,amsmath,amssymb,prb,superscriptaddress]{revtex4-1}
\usepackage{subfig}
\usepackage{float}
\usepackage{graphicx}
\usepackage{amssymb}
\usepackage{amsmath}
\usepackage{enumitem}
\usepackage{color}
\usepackage{textcomp}
\usepackage[symbol]{footmisc}

\begin{document}

\title{Efficient nitrogen-vacancy centers' fluorescence excitation and collection from micrometer-sized diamond by a tapered optical fiber}

\author{Dewen Duan}
\email{dduan@gwdg.de}
\affiliation{Max-Planck Research Group Nanoscale Spin Imaging, Max Planck Institute for Biophysical Chemistry,Am Fassberg 11,G\"{o}ttingen, 37077, Germany}
\author{Guanxiang Du}
\affiliation{Max-Planck Research Group Nanoscale Spin Imaging, Max Planck Institute for Biophysical Chemistry,Am Fassberg 11,G\"{o}ttingen, 37077, Germany}
\altaffiliation{Current address:School of Telecommunication and Information Engineering,Nanjing University of Posts and Telecommunications, Nanjing,China}
\author{Vinaya Kumar Kavatamane}
\affiliation{Max-Planck Research Group Nanoscale Spin Imaging, Max Planck Institute for Biophysical Chemistry,Am Fassberg 11,G\"{o}ttingen, 37077, Germany}
\author{Sri Ranjini Arumugam}
\affiliation{Max-Planck Research Group Nanoscale Spin Imaging, Max Planck Institute for Biophysical Chemistry,Am Fassberg 11,G\"{o}ttingen, 37077, Germany}
\author{Yan-Kai Tzeng}
\affiliation{Institute of Atomic and Molecular Sciences, Academia Sinica, Taipei 106, Taiwan}
\affiliation{Current adress: Department of Physics, Stanford University, Stanford, California 94305, USA}
\author{Huan-Cheng Chang}
\affiliation{Institute of Atomic and Molecular Sciences, Academia Sinica, Taipei 106, Taiwan}
\author{Gopalakrishnan Balasubramanian}
\email{gbalasu@gwdg.de}
\affiliation{Max-Planck Research Group Nanoscale Spin Imaging, Max Planck Institute for Biophysical Chemistry,Am Fassberg 11,G\"{o}ttingen, 37077, Germany}
\affiliation{Center Nanoscale Microscopy and Molecular Physiology of the Brain (CNMPB), G\"{o}ttingen, 37077, Germany}

\begin{abstract}

Efficiently excite nitrogen-vacancy (NV) centers in diamond and collect their fluorescence significantly benefit the fiber-optic-based NV sensors. Here, using a tapered optical fiber (TOF) tip, we significantly improve the efficiency of the laser excitation and fluorescence collection of the NV, thus enhance the sensitivity of the fiber-optic based micron-sized diamond magnetic sensor. Numerical calculation shows that the TOF tip delivers a high numerical aperture (NA) and has a high fluorescence excitation and collection efficiency. Experiments demonstrate that using such TOF tip can obtain up to over 7-fold the fluorescence excitation efficiency and over15-fold the fluorescence collection efficiency of a flat-ended (non-TOF) fiber. Such fluorescence collection enhances the sensitivity of the optical fiber-based diamond NV magnetometer, thus extending its potential application region. \\

\end{abstract}

\maketitle

\section{Introduction}
Negatively charged Nitrogen-vacancy (NV) color centers in diamond, as a remarkable solid-state quantum optical systems, has a high optical stability and a long spin lifetime at room temperature \cite{Jelezko2006, Doherty2013, Childress2006, Balasubramanian2009}, is an attractive candidate for biomarking/imaging\cite{Wu2013, Han2010}, magnetic field sensing\cite{Balasubramanian2008, Maze2008,Taylor2008, Rondin2014},  electric field  monitoring\cite{Dolde2014} and temperature measuring\cite{Neumann2013, Kucsko2013, Toyli2013}.  In most of these applications, the spin state of the NV center is initialized and detected by light guided through optical components such as objectives, lenses, and dichroic mirrors. The bulky size and poor flexibility of these optical components limit its portability and using-environment compatibility, thus the applications of the diamond NV-based sensor. Using an optical fiber to replace these bulk components will greatly extend the potential application areas of the NV-based sensor. 

In recent years, some research groups have integrated NV centers in an optical fiber to explore its potential application region. For example, Henderson \textit{et al.} and Ruan \textit{et al.}  have embedded nanodiamonds into tellurite glass to fabricate optical fibers with nanodiamonds inside fibers\cite{Henderson2011, Ruan2015}. Rabeau \textit{et al.} have deposited a few diamonds onto a fiber end face\cite{Rabeau2005}.  Schr\"oder \textit{et al.}  have placed a nanodiamond containing a single NV center on the end face center of a microstructure fiber to form an in-fiber single-photon source\cite{Schroede2011}. Mayer \textit{et al.} have coupled a diamond containing NV centers to a single-mode photonic crystal fiber \cite{Mayer2016}. Liebermeister \textit{et al.} have used a tapered optical nanofiber to collect the fluorescence from a single NV center in a nanodiamond on the surface of the nanofiber\cite{Liebermeister2014}.  Patel \textit{et al.} have fabricated an optical fiber nanotaper and collected over 16\% of fluorescence from a single NV center in a nanodiamond wire\cite{Patel2016}. Similarly, Burek \textit{et al.} also used a single mode fiber tip and collected over 90\% fluorescence from a single color center (Silicon-Vacancy) in a photonic crystal cavity fabricated in a diamond  waveguide\cite{Burek2017}. Fedotov\textit{et al.} have glued a  \( 200\ \mu \)m and  \( 30\ \mu \)m sized high NV center density diamond onto the flat end of a multimode fiber and a high numerical aperture photonic crystal optical fiber respectively; they excited the NV centers and collected its fluorescence response through the same fiber in their sensor systems\cite{Fedotov2014s, Fedotov2014o, Fedotov2014a, Fedotov2016o}. In NV spins' metrological applications, the collected  NV's fluorescence of each measurement is directly linked to its minimum detectable variation; therefore, enhanced collection efficiency directly provides a better sensitivity \cite{Pham2011}. Among the aforementioned methods, a tapered nano-fiber is an efficient way to collect the diamond NV-center fluorescence. However, the enhancement does not scale well for micron-sized diamond. On the other hand, attaching the diamond to the flat end of an optical fiber is easy to achieve, but it offers only a limited excitation and collection efficiency significantly lower than using a high numerical aperture (NA) objective.

Here, we demonstrate a method to efficiently excite NV centers in micrometer-sized diamond crystal and collect its emission fluorescence through the same fiber. Our method is based on using a quasi-adiabatic tapered optical fiber (TOF) tip made from a multimode optical fiber (shown in Fig.\ref{fig: Figure1}(a)) to concentrate the excitation laser on to the diamond, at the same time, to collect the fluorescence from the diamond with a high efficiency. The TOF tip is equivalent to a high NA objective on the optical fiber. Numerical calculations show that such TOF tip delivers a ultra-high NA larger than 1 and has a high fluorescence excitation and collection efficiency.  Experiments demonstrate that when exciting and collecting fluorescence from a $\sim$5-\(\mu \)m diamond, using such TOF-fiber tip can achieve over 7-fold the fluorescence excitation efficiency and up to 15-fold the fluorescence collection efficiency of a flat-ended (non-TOF) fiber (table.\ref{tbl: comparison}). Comparing Optically Detected Magnetic Resonance (ODMR) scanning experiments demonstrated that this TOF tip can boost the magnetic field sensitivity of a micro-sized NV sensor. In a direct current magnetic field measuring,  a TOF based $\sim$7.9-\(\mu \)m diamond NV-magnetometer achieve 1/28 the sensitivity value of a non-TOF based $\sim$11.4-\(\mu \)m diamond NV-magnetometer, reach a sensitivity of 180nT/$\sqrt{Hz}$. 

\section{Principles and methods}

\begin{figure}[ht]

\includegraphics[width=3.30in]{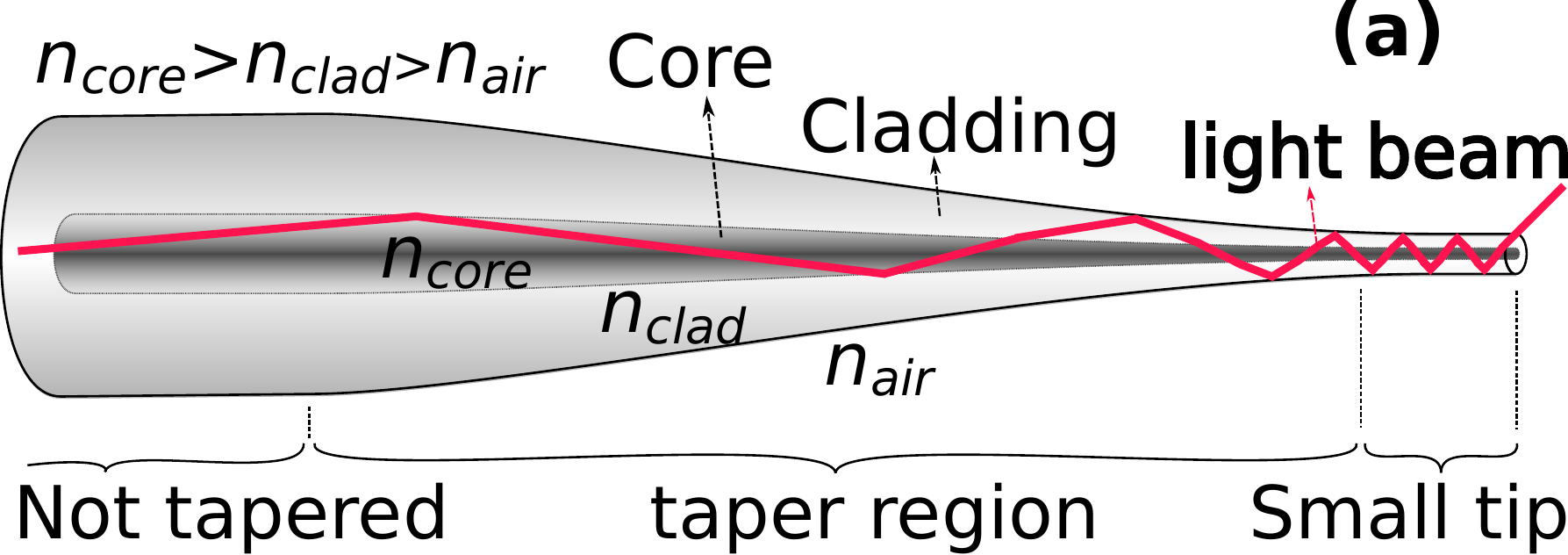} 
 
\includegraphics[width=3.30in]{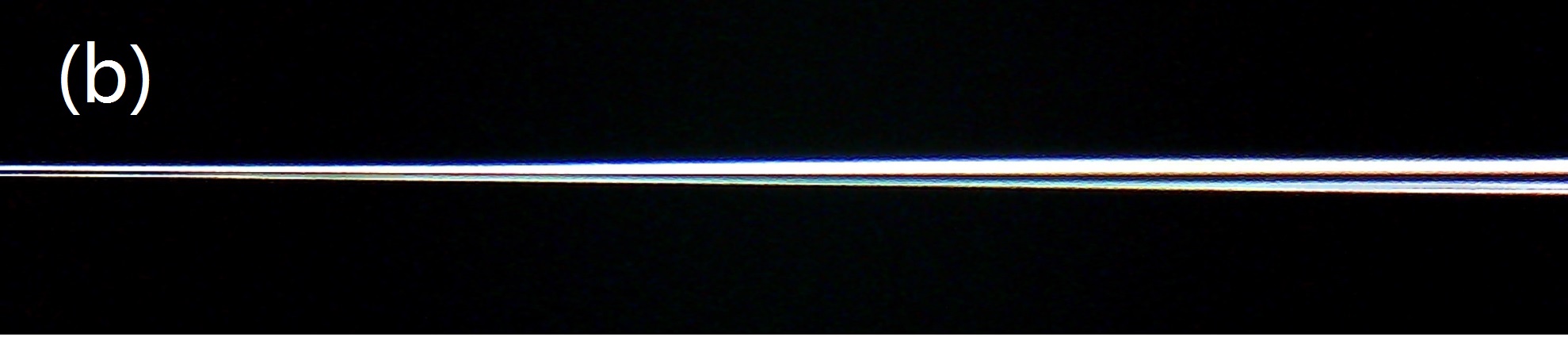}

\includegraphics[width=3.30in]{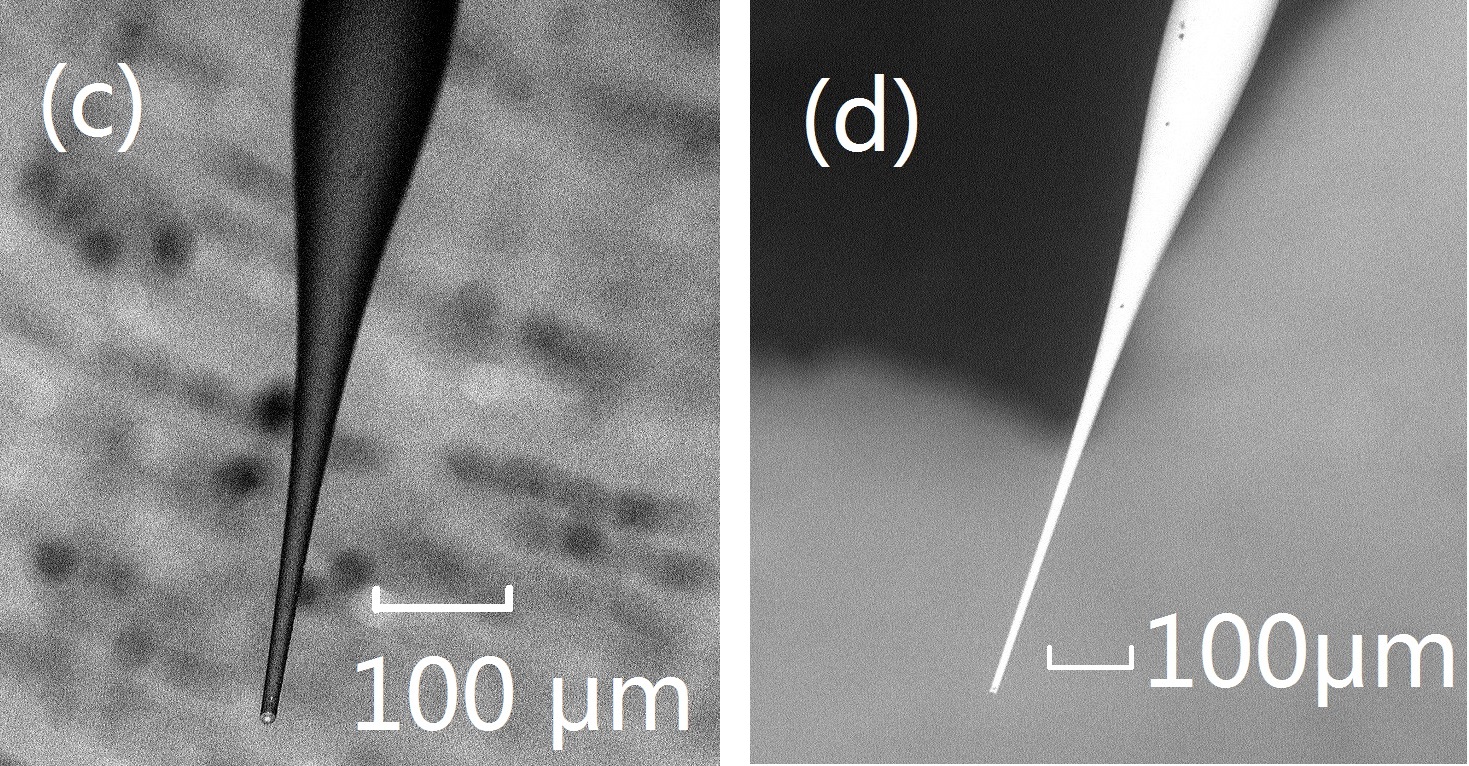}
   \caption{The tapered optical fiber (TOF) tip. (a) Schematic diagram of the light ray trace that resulting in the high NA for the multimode optical fiber based TOF tip. (b) Optical microscope image of part of a fabricated long TOF tip's tapered region (total TOF length is $\sim$15 mm, the photo is taken with 50 times magnification).  (c) the side view and (d) the 45-degree view of another fabricated short TOF tip with a length of $\sim$1 mm in a scanning electron microscope.} 
\label{fig: Figure1}
\end{figure}

\begin{figure}[ht]
    \centering
    \includegraphics[width=3.33in]{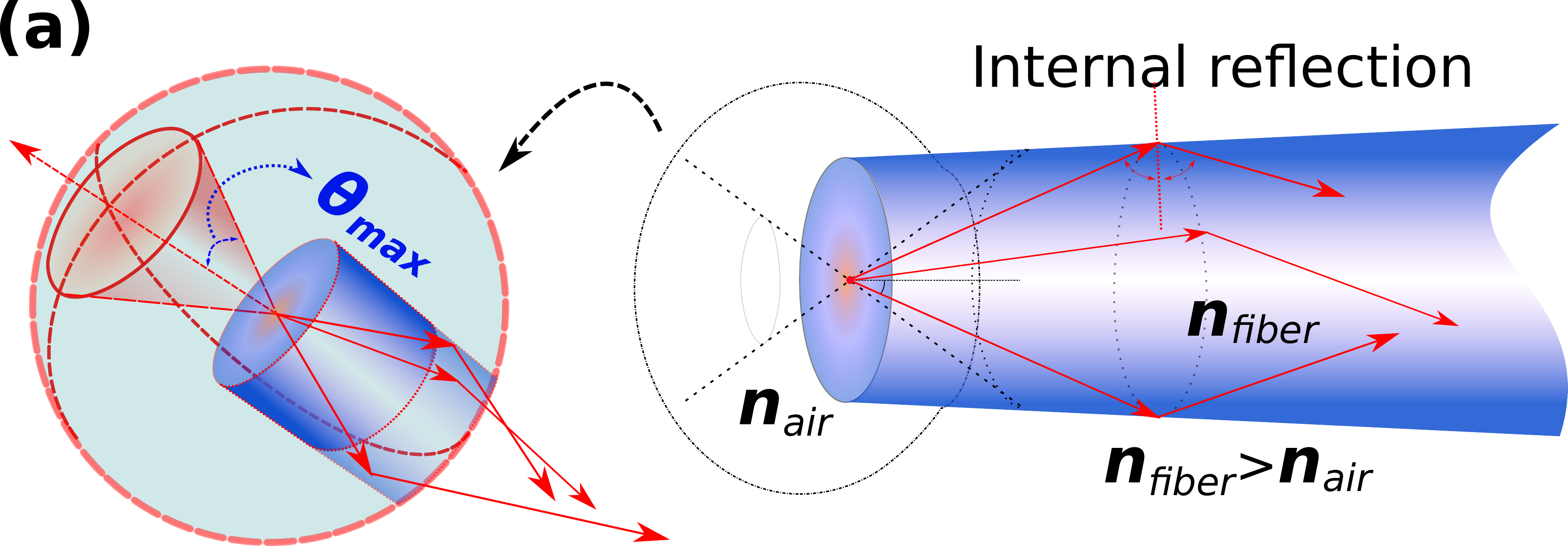} \\
 
     \includegraphics[width=3.33in]{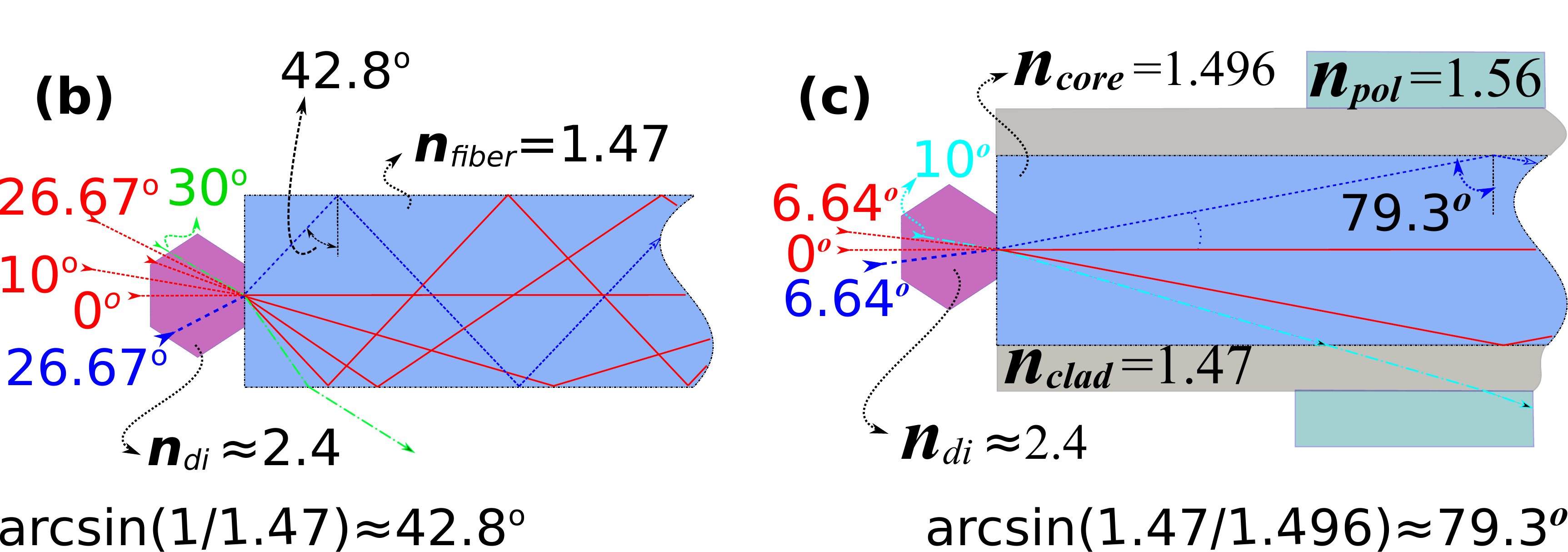} \\
     \caption{The collection efficiency calculation schematic diagram of the TOF tip and flat fiber end. (a) is the fiber light collection solid angle.  (b) and (c) are fluorescence acceptance angles of the TOF tip and bare flat fiber end when collecting light from a contacted diamond respectively.}
\label{fig: Figure2}
\end{figure}

\begin{figure}[ht]
    \centering
    \includegraphics[width=3.33in]{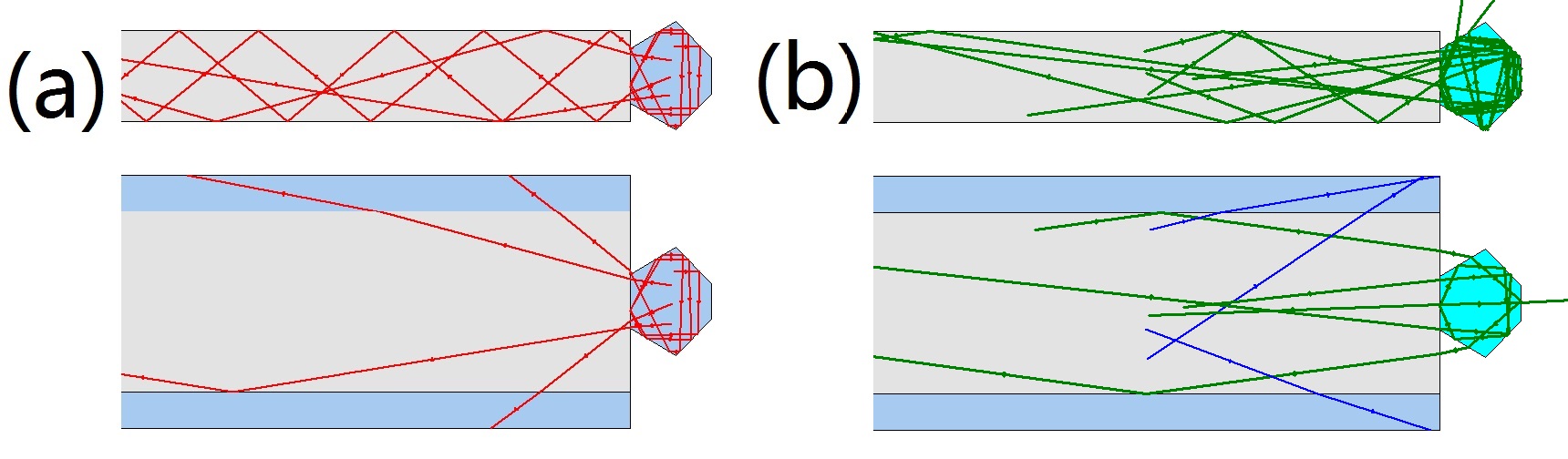}
    \caption{Simulation of multiple reflections in (a) NV fluorescence collection and in (b) NV excitation for both TOF tip (top) and bare flat fiber end (bottom). In (a) the original fluorescence rays are directed in $ 2^{\circ}$, $171^{\circ}$, $203^{\circ}$ and $186^{\circ}$ from same points/positions of the diamond in both the diamonds respectively. At the bottom of (b), the rays (blue) did not enter the diamond are not exist rays just for comparison. (The simulation is carried out by free software Optgeo: http://jeanmarie.biansan.free.fr/optgeo.html.)}
\label{fig: Figure3}
\end{figure}

The TOF concept stems from the single-mode fiber TOF called adiabatic taper.  Decreasing the diameter shrinkage ratio of an optical fiber taper can reduce the light transmission loss of the taper region; when the shrinkage ratio reached a certain value, the transmission loss can be almost ignored\cite{Love1991, Black1991}. Through multiple total internal reflections, the diameter slowly shrinking TOF condenses the 532-nm laser from the multimode fiber core into the thin TOF tip, at the same time, converts the fluorescence entered the thin TOF tip within its maximum acceptance angle into the core of the multimode fiber (Fig.\ref{fig: Figure1}(a) and Fig.\ref{fig: Figure2}.(a)). This nearly adiabatic process enhances the NV centers' fluorescence excitation and collection. Calculated in an ideal geometry ray optics, the TOF tip's fluorescence collection efficiency can reach approximately  $\sim$16 times of a bare flat fiber end ( Fig.\ref{fig: Figure2} (b) and Fig.\ref{fig: Figure2} (c)) (when assuming the diamond's refractive index is $ n_{di}\approx 2.4$). In experiments, using a long TOF ($\sim$ \( 7.4\ \mu \)m in tip diameter and $\sim$ 15 mm in length), we achieved 15.03 times fluorescence collection efficiency of a bare flat fiber end in collecting fluorescence from the same  \(5\ \mu \)m diamond (table.\ref{tbl: comparison} ). A simulation example of multiple reflections comparison of both fluorescence collection and excitation between TOF tip and bare fiber end is shown in Fig.\ref{fig: Figure3}.

\subsection{TOF tip and diamonds preparing}

The optical fiber we used is a graded index multimode fiber (GIF625, Thorlabs. Core diameter: \( 62.5\ \mu \)m, cladding diameter:\( 125\ \mu \)m, NA : $0.275\pm0.015$ @ 850 nm). The TOF tip was fabricated by first melting and stretching the fiber and then cutting the taper at its waist. By adjusting the heating and stretching force, the shape of the taper can be modified. More details about available tapering techniques to control the taper shapes and taper optical properties have been previously published in the literature\cite{Birks1992, Stiebeiner2010, Pricking2010, Latifi2012, Harun2013}. We fabricated two kinds of TOF tips: long TOF tip with a small taper's diameter shrinkage ratio tapered under flame heating (Fig.\ref{fig: Figure1}(b), over $\sim{10} $ mm long), and short TOF tip with a large taper's diameter shrinkage ratio tapered under arc charging heating of an optical fusion splicer (Fig.\ref{fig: Figure1}  (c) and (d), less than $\sim$1.5 mm long). 

The diamonds were prepared by irradiating the high-temperature high-pressure (HPHT) synthetic monocrystalline diamond powders with 3-MeV H+ to form the vacancies in the diamond. Subsequently annealed the diamond samples in vacuum at 800 $^{o}$C for approximately two hours to concentrate the nitrogen and vacancies\cite{Wu2013}.

\subsection{TOF tip collection efficiency calculation}

An acceptance angle $\theta_{max}$ related solid angle (the spherical surface area of the angle related ball crown divided by the square of the spherical radius) divided by $4\pi$ is its collection efficiency (Fig.\ref{fig: Figure2}(a)). Taking the diamond attached on the fiber tip surface as a thin sheet point sources ensemble, the fiber tip's collection efficiency is
 
\begin{equation}
   \eta=\frac{2\pi r^{2}[1-cos( \theta_{max})]/r^{2}} {4\pi}=\frac{1-cos( \theta_{max})}{2}
   \label{eqn:1}
\end{equation}

NA is defined as $ NA=(n_{core}^2- n_{clad}^2)/2=0.275$ for the used optical fiber; ($ n_{core}$ is the refractive index of the fiber core and $ n_{clad}$ is the refractive index of the fiber cladding). The group refractive index of GIF625 fiber is 1.496 (@ 850 nm, the GIF625 fiber data sheet); then the cladding refractive index of the fiber  $ n_{clad}\approx1.47$. Ignoring the refractive index distribution in the cross-section of the TOF tip and treating it as the core with air surrounding it as the cladding. The maximum acceptance angle of the TOF tip is calculated according to Fig.\ref{fig: Figure2}(b). 

\begin{equation}
   \theta_{max} = arcsin[\frac{sin(\pi/2-arcsin(n_{clad}/n_{core})} {n_{di}}\times n_{core}]  \label{eqn:2}
\end{equation}

Taking $n_{di}\approx2.4$ as the refractive index of diamond, $n_{fiber} = n_{clad}\approx1.47$  as the TOF tip's refractive index and $n_{clad}=n_{air}\approx1$, a maximum acceptance angle of $\theta_{max} = 26.676^{\circ}$ is obtained for the TOF tip (Fig.\ref{fig: Figure2}(b). If we take $n_{fiber} = n_{core}\approx1.496$  as the TOF tip refractive index, $\theta_{max} = 27.620^{\circ}$). 
Using equation (\ref{eqn:1}), the TOF's collection efficiency can reach $\eta_{d}\approx5.322\%$ (If we take $\theta_{max} = 27.620^{\circ}$, $\eta_{d}\approx5.698\%$) in collecting fluorescence from an attached diamond. For the flat bare fiber end, using equation (\ref{eqn:2}), taking $n_{core}=1.496 $ and $n_{clad}=1.47 $, the calculated $\theta_{max}\approx6.644^{\circ}$ and the collection efficiency $\eta_{0.275d}\approx 0.336\%$ for accepting fluorescence from a contacted diamond (Fig.\ref{fig: Figure2} (d)).  The TOF tip's fluorescence collection efficiency is $\sim16$ times of a bare flat fiber end ($\eta_d/\eta_{0.275d}\approx5.322/0.336\approx16$; if we take $\eta_{d}\approx5.698\%$, $\eta_d/\eta_{0.275d}\approx5.698/0.336\approx17$ ).  Taking $\theta_{max} = 26.676^{\circ}$ the calculated NA = $n_{di} \times sin(\theta_{max})$ of the TOF tip is $NA\approx 1.077$ (if take  $\theta_{max} = 27.620^{\circ}$, $NA\approx 1.113$).

As the strongest fluorescence of a diamond NV center is 630-780 nm instead of 850 nm at which the fiber parameters are given, and there are internal multiple reflections in the diamond, the exact collection efficiency values may deviate from the calculated values. However, these simple calculations still illustrate that the TOF tip will enhance the fluorescence collection efficiency significantly. 

Since the not tapered graded-index multimode fiber can partly concentrate the laser light in the central core of the fiber, the bare flat fiber end has a relatively high excitation efficiency in illumination a micrometer-size diamond attached at its end center. The TOF can further concentrate laser light on to its tip and enhance its fluorescence excitation efficiency.  But its enhancing value will be relatively smaller than its collection efficiency enhancing.

\section{Results and discussion}

\begin{figure}[ht]
    \centering
    \includegraphics[width=3.0in]{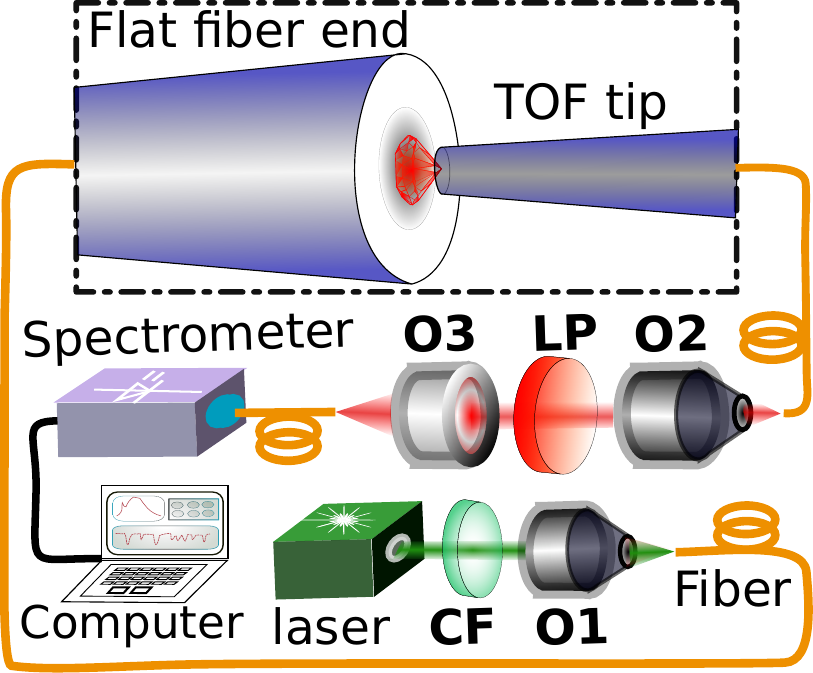}
 \caption{TOF tip's fluorescence collection and excitation efficiency enhancement assessing setup. O1, O2, and O3 are objectives; CF is a 512-nm clear-up filter and LP is a 615-nm long-pass filter. The diamond is attached (by UV-curing glue) to the center of a bare fiber tip end (enlarged in the dashed box). In the fluorescence collection efficiency measuring, the diamond is illuminated under constant laser intensity through the bare fiber tip, the fluorescence is collected by the tested TOF tip or flat fiber end. In the fluorescence excitation efficiency measuring (fiber tip allying need to be reversed in the enlarged dashed box), the fluorescences are collected by the bare fiber tip and the diamond is illuminated through the bare fiber end or the tested TOF tip with constant laser intensity coupled into them.}
\label{fig: Figure4}
\end{figure}

\begin{figure}[ht]
\includegraphics[width=3.0in]{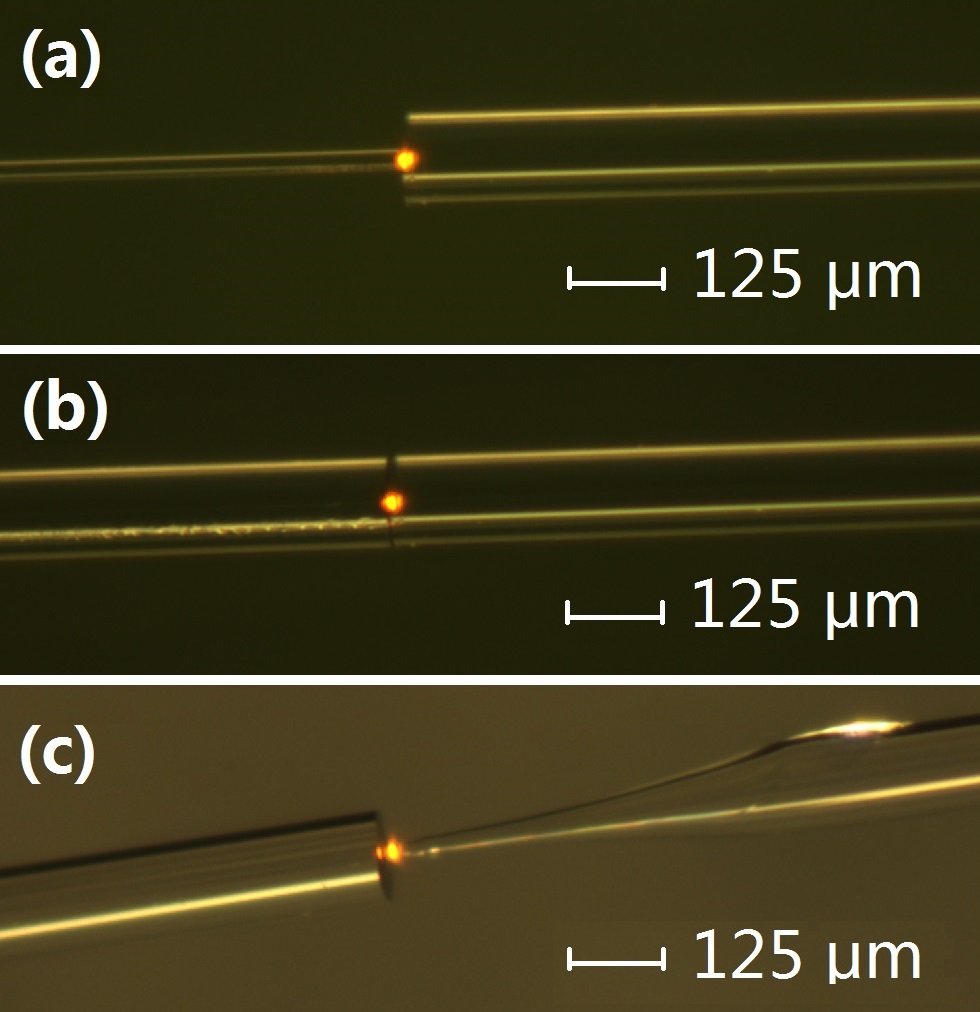} 
 \caption{Microscope images of fluorescence excitation and collection efficiency measuring sets of (a) a TOF tip, (b) a flat fiber end and (c) a short TOF tip.}
\label{fig: Figure5}
\end{figure}

The TOF fluorescence collection and excitation efficiency enhancement assessing experiment setup schematic and microscope images are shown in Fig.\ref{fig: Figure4} and Fig.\ref{fig: Figure5}. The results are shown in table.\ref{tbl: comparison}. We obtained the fluorescence collection efficiency enhancing value "collect" by comparing the intensities of fluorescence collected from the tested TOF tip and bare fiber end respectively when the diamond is attached on a bare fiber tip and irradiated with constant laser intensity through the bare fiber tip (Fig.\ref{fig: Figure4}). In this way, the diamond emits constant fluorescence out, the collected fluorescence intensity will only depend on the tested collection tip's collection ability. Similarly, we obtained the fluorescence excitation efficiency enhancing value "excite" by comparing the collected fluorescence intensities through the bare fiber tip attached to the diamond when the diamond is irradiated by the tested TOF tip and flat fiber end respectively under constant laser intensity coupled into them (fiber tip allying need to be reversed in the enlarged dashed box of Fig.\ref{fig: Figure4}). In this way, the collection  bare fiber tip's fluorescence collection ability is fixed, and the coupled laser intensity into the TOF tip and flat fiber end is constant, the diamond emits fluorescence intensity, so as the collected fluorescence intensity, will only depend on the tested TOF tip and flat fiber end's excitation ability (concentrate the laser light into the diamond). In all experiments, both the tested TOF tip and the bare fiber end are dipped with a tiny amount of immersion oil (Fluka 51786, refractive index: 1.512 - 1.522) to enhance its contact to the diamond.  The lengths of long TOF (long taper in table.\ref{tbl: comparison}) and short TOF (short taper in table.\ref{tbl: comparison}) are $\sim$15mm and $\sim$1.1 mm respectively. 

\begin{table*}
  \caption{\label{tbl: comparison} Excitation and collection enhancement}
    \begin{tabular}{cccccccc}
    \hline
      diamond size &  &  \multicolumn{2}{c}{\( 5\ \mu \)m}   &  \multicolumn{2}{c}{\( 11.7\ \mu \)m} &  \multicolumn{2}{c}{\( 50\ \mu \)m} \\
    
      &tip size & collect\textsuperscript{\emph{a}}  &excite\textsuperscript{\emph{b}} & collect\textsuperscript{\emph{a}}  & excite\textsuperscript{\emph{b}} & collect\textsuperscript{\emph{a}} & excite\textsuperscript{\emph{b}} \\
       \hline
     long  taper1& \( 7.4\ \mu \)m&15.03  & 7.68 & 6.1  & 2.04 & 1.35 & 0.97 \\
     long taper2&\( 9.6\ \mu \)m &13.13  & 7.84 & 10.79  & 2.13 & 1.85 & 1.11 \\
     long taper3& \(17.6\ \mu \)m&5.35  & 5.73 & 8.51  & 1.93 & 2.11 & 1.12 \\
     short taper& \(12.6\ \mu \)m&--  & -- & 5.64  & 2.42 &--  &-- \\
     bare fiber end& \(125\ \mu \)m&1  & 1 & 1  & 1 & 1 & 1 \\   
    \hline
    \end{tabular}

   \textsuperscript{\emph{a}} ratio of the collected fluorescence intensity of the TOF tip and the bare fiber end;\\
  \textsuperscript{\emph{b}} ratio of the excited fluorescence intensity of the TOF tip and the bare fiber end.
\end{table*}

\begin{table*}
  \caption{Obtain 262 nW fluoresence needed excitation laser power}
  \label{tbl: excitation}
\begin{tabular}{llllllllll}
    \hline          
     diamond size && fiber end type && tip size && taper length && require laser power \textsuperscript{\emph{a}}\\
        \hline
     \(12.5\ \mu \)m && long  taper&& \( 12.8\ \mu \)m & &15mm  && \(151\ \mu \)W \\
     \(12.5\ \mu \)m && bare fiber end&&\( 125\ \mu \)m & & --  & &4.57mW  \\
     \(156\ \mu \)m && bare fiber end &&\( 125\ \mu \)m &  &--  & &1.52mW  \\
     \(7.9\ \mu \)m &&  long taper&& \(12.2\ \mu \)m      & &12mm  &&  \(333\ \mu \)W \\
     \(11.3\ \mu \)m && long taper && \(13.3\ \mu \)m &&10mm  &&  \(202\ \mu \)W \\
     \(11.7\ \mu \)m && short taper& &\(16.8\ \mu \)m &&1 mm  && \(846\ \mu \)W \\
     \(11.7\ \mu\)m && bare fiber end& &\(125\ \mu \)m  && --  && 8.07 mW \\   
    \hline
  \end{tabular}
   
   \textsuperscript{\emph{a}} The laser power coupled into the optical fiber attached to the diamond\\
\end{table*}

Table.\ref{tbl: comparison} indicate that the long TOF tip can obtain a maximum value of 15.03 times the fluorescence collection efficiency of a bare flat fiber end, it agrees to our calculated 16 times based on geometry ray optics. This is because our TOF and diamond have a dimension scale larger than ten times the wavelengths of the laser or fluorescence, fit the geometry ray optics using condition. Table.\ref{tbl: comparison} also shows that the fluorescence collection efficiency is sensitive to the relative size of the TOF tip and diamond; generally, a smaller tip size TOF will have a better collection efficiency for a smaller diamond, when the diamond size is set, using a TOF tip with a tip size close to the diamond can collect more fluorescence. While for fluorescence excitation, if the TOF tip size is too small, it will reduce the excitation efficiency. Using a TOF tip with a tip size of $\sim$\(10\ \mu \)m can get the maximum excitation efficiency. 

\begin{figure}[ht]
\includegraphics[width=3.0in]{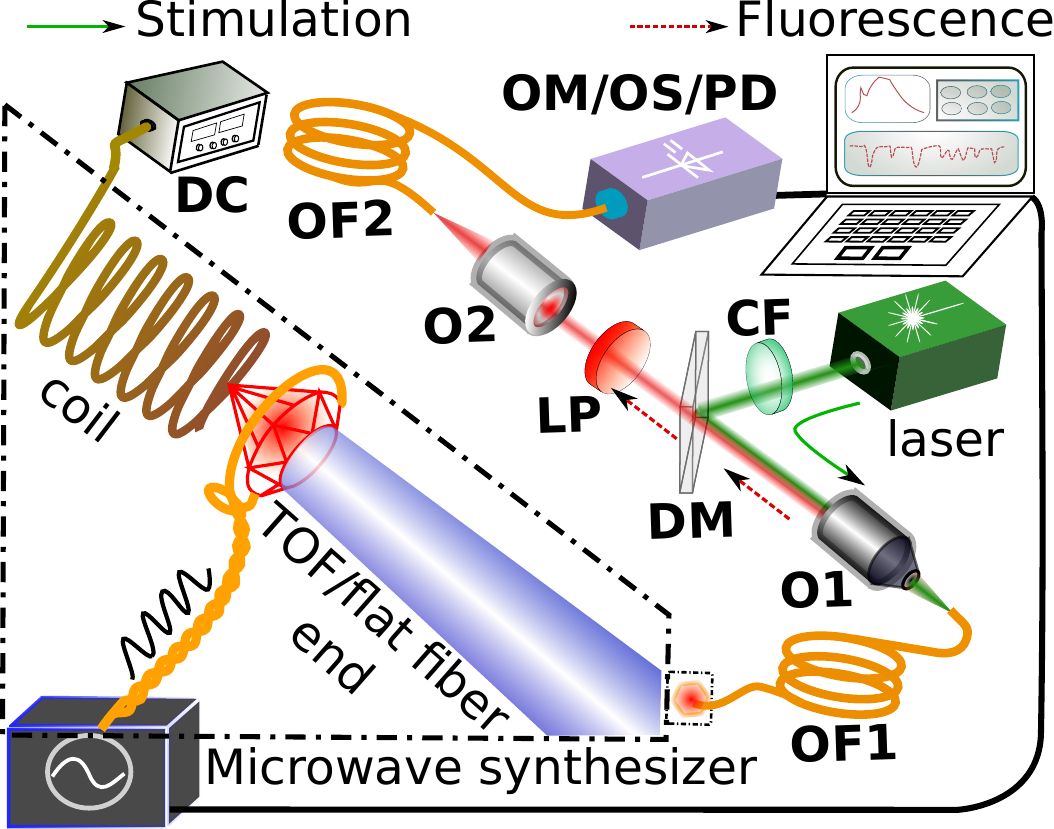} 
 \caption{Fiber-based diamond NV center system setup. The 532-nm laser (solid arrow), after passing the 532-nm clean-up filter CF and the dichroic mirror DM, is coupled by the objective O1 into the optical fiber OF1 tip where the diamond is attached (dashed triangular box). The fluorescence collected by the TOF/Flat fiber end cleared by 615-nm long-pass filter LP  (dashed arrow) and coupled by objective O2 into optical fiber OF2 which is connected to the optical power meter/optical spectrometer/photodiode OM/OS/PD. The DC power supply is used to apply DC magnetic field to the diamond}
\label{fig: Figure6}
\end{figure}

\begin{figure}[ht]
\includegraphics[width=3.33in]{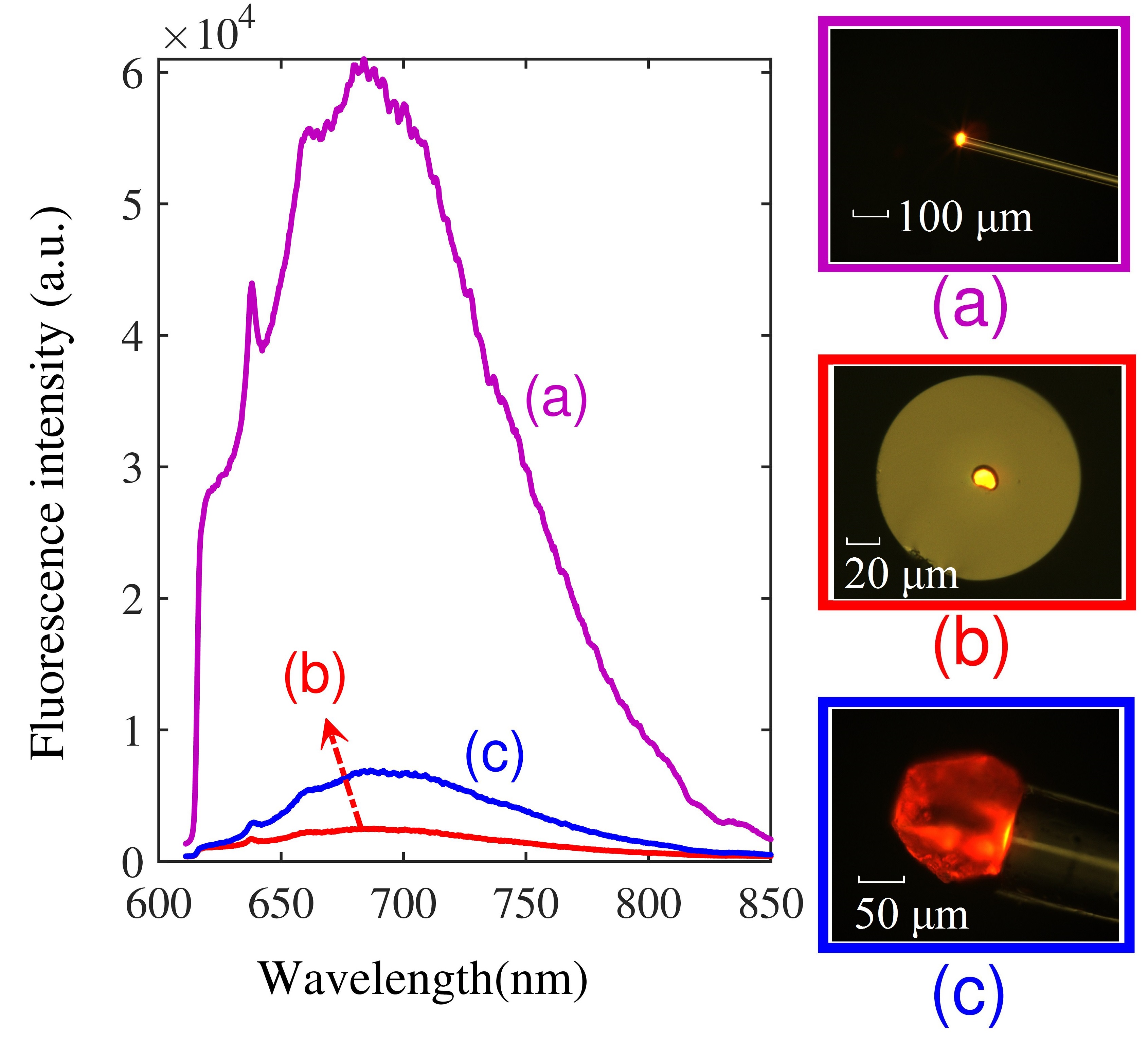} 
 \caption{Comparison of measured fluorescence spectra. In all cases, the excitation laser power launched into the fiber is $\sim$151 \(\mu\)W. (a) is the spectrum of a $\sim$12.5-\(\mu \)m diamond on a TOF tip, (b) is the spectrum of a $\sim$12.5-\(\mu \)m diamond on a cleaved flat fiber end, and (c) is the spectrum of a $\sim$156-\(\mu \)m diamond on a cleaved flat fiber end. The inset shows microscope images of these samples. The length of the TOF is $\sim$15 mm, and its tip diameter is  $\sim$12.8\(\ \mu \)m. The fiber diameter is $\sim$125\(\ \mu \)m.}
\label{fig: Figure7}
\end{figure}

We positioned another $\sim$\(12.5\ \mu \)m NV enriched diamond crystal on a  long TOF (tip diameter: $\sim$\(12.8\ \mu \)m, length: $\sim$15 mm. Inset (a)  of Fig.\ref{fig: Figure7}). We used the setup shown in Fig.\ref{fig: Figure6} and coupled $\sim$\( 151\ \mu \)W 532-nm laser light into the fiber to illuminate the diamond attached on the TOF tip and collected $\sim$262 nW fluorescence through the same TOF tip (the power meter wavelength was set at 635 nm).  Subsequently, we cut off the TOF tip and transferred the same diamond onto the center of the flat fiber end (inset (b) of Fig.\ref{fig: Figure7}). We find that to obtain $\sim$262 nW fluorescence, approximately 4.57 mW of laser power ($\sim$30 times more than the TOF tip needed) is required to be coupled into the same fiber. Fig.\ref{fig: Figure7} shows the collected fluorescence spectra of the $\sim$\(12.5\ \mu \)m NV enriched diamond attached on a long TOF tip, flat fiber end, and a $\sim$\(156\ \mu \)m NV enriched diamond attached to a flat fiber end (inset (c) of Fig.\ref{fig: Figure7}) with UV-curing glue when coupling $\sim$\( 151\ \mu \)W of the 532-nm laser into the optical fiber.  We also positioned some other size diamond on to long TOF tip, short TOF tip, and bare flat fiber end respectively. Table.\ref{tbl: excitation} shows the required laser power coupling into the optical fiber to obtain 262 nW fluorescence through the same fiber for different fiber tips. It is clear, a TOF with short taper length exhibits a larger loss in transforming the light between the tapered tip and the not tapered fiber region due to the large fiber diameter shrink ratio cannot meet the adiabatic criteria; as a result, a short-TOF has a lower fluorescence excitation and collection efficiency than that of the long-TOF. However, a shorter TOF tip can provide a better mechanical stability than the longer TOF, which may benefit certain applications that require high TOF tip robustness. 

\begin{figure}[ht]
\centering
      \includegraphics[width=3.0in]{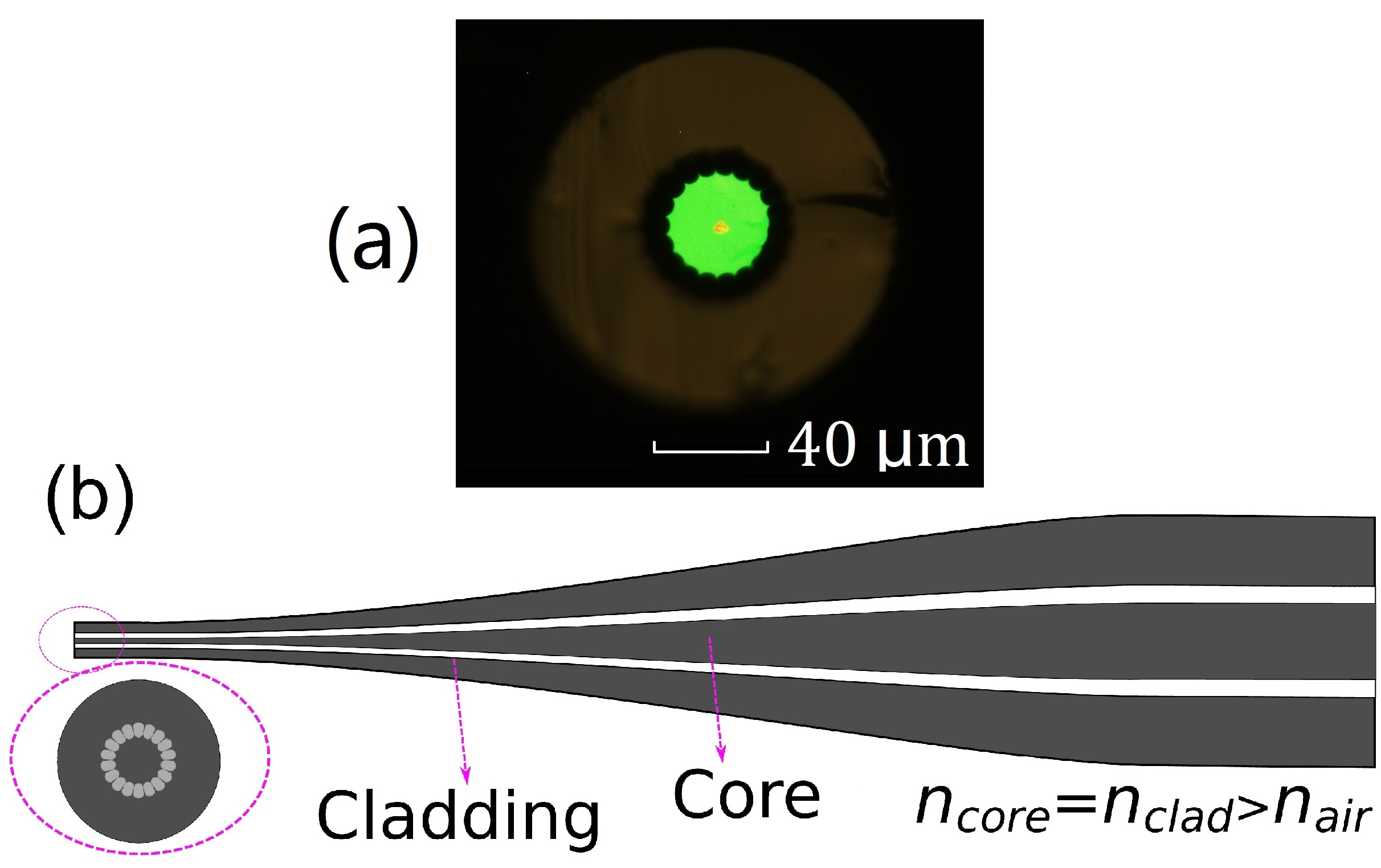}\\
      \caption{ (a) Microscope images of a $\sim$5.2\(\ \mu \)m diamond on the center of the cleaved flat fiber end of the Ultra-high NA photonic microsctructure fiber. (b) Schematic diagram of the tapered ultra-high NA photonic microstructure fiber.}
\label{fig: Figure8}
\end{figure}

For comparison, we used the Fig.\ref{fig: Figure6} setup and investigated the possibility of using an ultra-high NA photonic microstructure fiber MM-HNA-35 (NKT Photonics, Denmark) in fluorescence excitation and collection from diamonds smaller than ten micrometers. Surprisingly, we find that when coupled in with equal laser power, the MM-HNA-35 fiber even collected less fluorescence than not tapered GIF625 fiber. We believe this is because, unlike the graded-index core GIF625 fiber, the 35 \(\mu \)m pure silica core of the MM-HNA-35 fiber cannot focus the laser light into its center (Fig.\ref{fig: Figure8}.(a)); thus, it is poor at excitation diamond smaller than its core.  When tapering the MM-HNA-35 fiber, the air isolation band between the core and cladding will partly keep ( Fig.\ref{fig: Figure8} (b) shows its schematic illustration), most of the fluorescence entered the cladding region (92\% of the tapered tip: $1-(35/125)^2\approx0.92 $) cannot be coupled into the core of the fiber; as a result, the tapering enhanced laser excitation cannot compensate for it decreased fluorescence collection. Thus it is not surprising to find out that the TOF form of the MM-HNA-35 tip even has collected less fluorescence than that of its un-tapered form when coupled in an equal laser power.  In principle, tapering the MM-HNA-35 fiber into TOF tip with tip's core equal to the used diamond or directly design a fiber with a core equal to the used diamond will solve this problem.
 
 In addition, this TOF tip can combine the micro-concave technique \cite{Duan2018} (locate the diamond attached on the TOF tip at the focal point of a matched micro-concave mirror facing against the tip) to further enhance the fluorescence excitation and collection efficiency. As the micro-concave mirror can focus back the fluorescence emitted in the opposite direction of the TOF in a smaller incidence angle and the TOF has a relatively large fluorescence acceptance angle, this combining can greatly improve the system's fluorescence collection.  Ideally, if the matched micro-concave mirror is in parabolic reflector shape, the total fluorescence collection efficiency will be around 50\% + TOF tip (without using of the micro-concave mirror) fluorescence collect efficiency.  If the TOF tip's fluorescence collection efficiency is 5\%, combine it with a matched micro parabolic mirror will reach a fluorescence collection efficiency of 55\%, and enhance10 times the fluorescence collection efficiency. In a simple experiment evaluation, for an unmatched relatively big micro-concave mirror, we obtained only $\sim$ 2.2 times the collected fluorescence efficiency of the TOF tip without using the micro-concave mirror. 
 
The excitation efficiency of NV-center on a TOF tip is the number of generated fluorescence photons divide the number of 532-nm photons coupled into the optical fiber. Taking the $\sim$12.5-\(\mu \)m diamond on the  $\sim$12.8-\(\mu \)m TOF tip for example. The excitation laser is $\sim$\(151\ \mu \)W and the collected fluorescence is $\sim$262 nW. Using a wavelength calibrated calculation, we obtained the collected fluorescence photon number and incidence laser photon number ratio $\sigma \approx1.864/1000$. Divided by the calculated fluorescence collection efficiency 5.3\%, an NV-center excitation efficiency of 3.52\% is obtained. 

\begin{figure}[ht]
\includegraphics[width=3.0in]{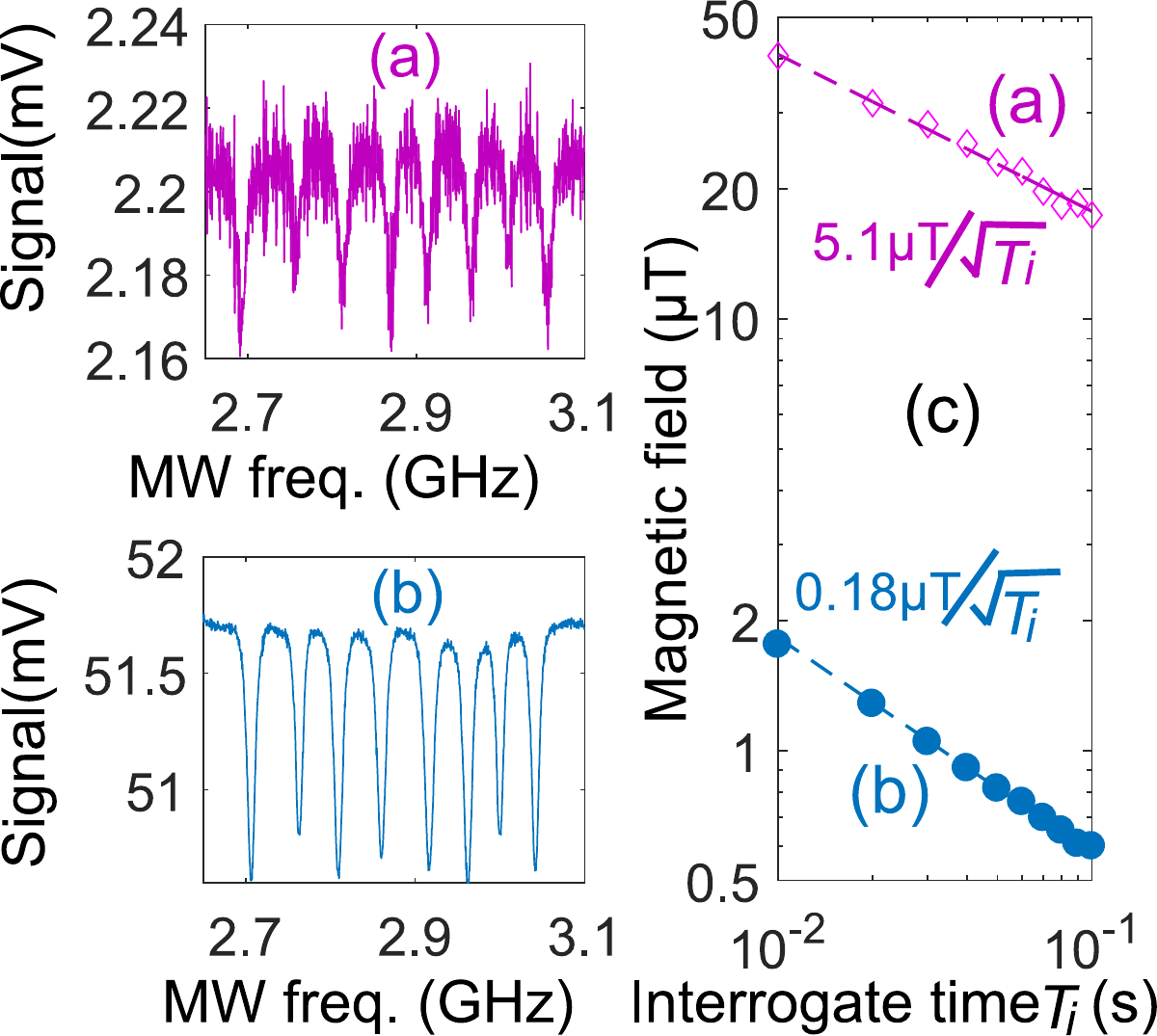} 
 \caption{(a) ODMR spectra of a $\sim$11.4-\(\mu \)m diamond UV-curing glue bonded on a flat fiber end center in a 9.6 mT magnetic field, and (b) the ODMR spectra of the UV-curing glue bonded$\sim$7.9-\(\mu \)m diamond on TOF tip in a 9.0 mT magnetic field, and the laser launched into the optical fiber for stimulation both diamonds are the same $\sim$36.5\(\ \mu \)W. The signals are read from a photodiode detector. (c) The Magnetic field sensitivity comparison of the two sensor heads (a) and (b). }
\label{fig: Figure9}
\end{figure}

\begin{figure}[ht]
\includegraphics[width=3.33in]{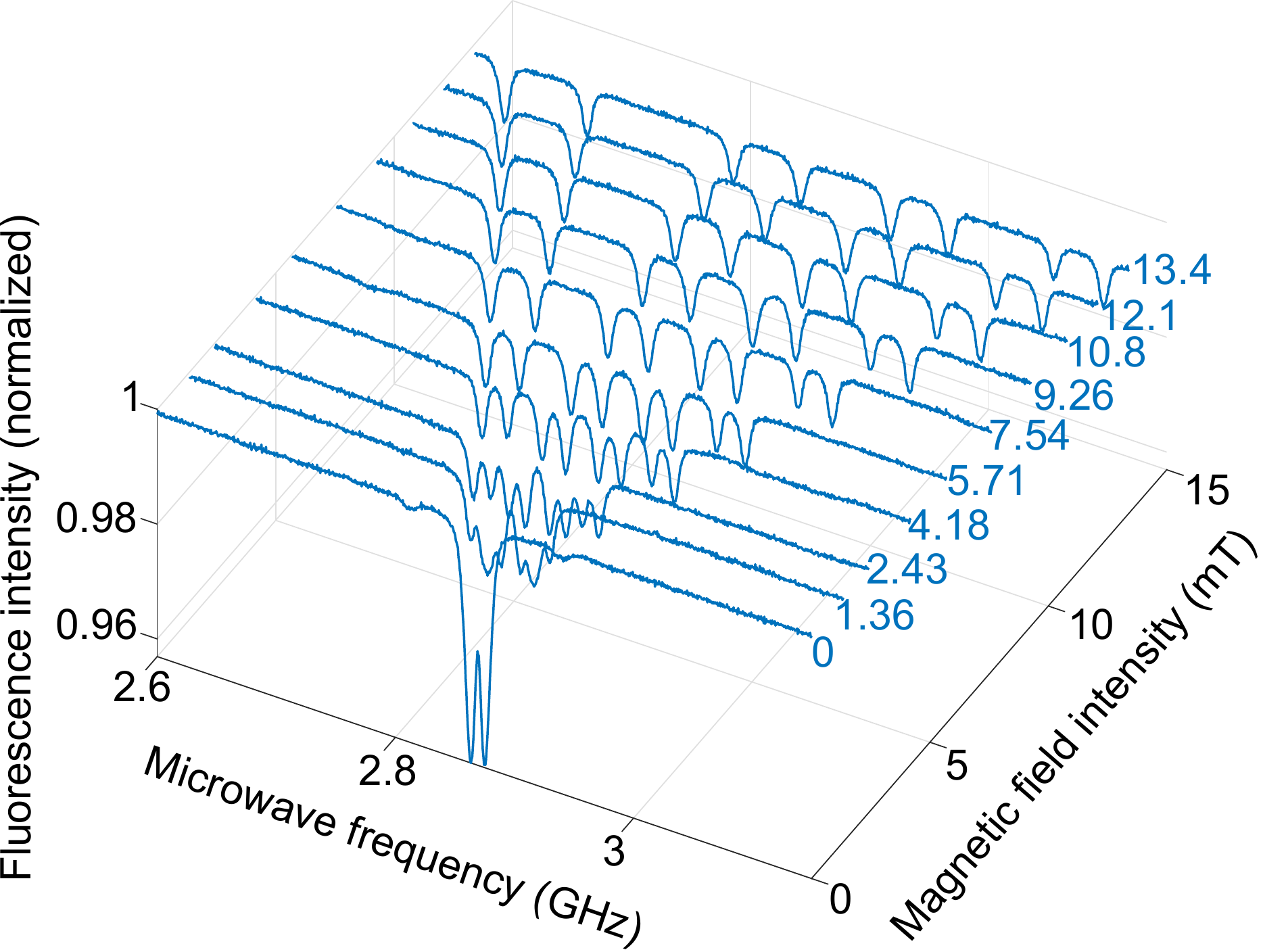} 
 \caption{ODMR spectra of the NV-center ensembles in a $\sim$11.3-\(\mu \)m diamond bonded on the tip of a $\sim$10-mm long TOF recorded for various magnetic field values (Fluorescence intensity have been normalized). The TOF tip' diameter is $\sim$13.3\(\ \mu\)m. The magnetic field was applied by the DC coil. The test was executed using the Fig.\ref{fig: Figure6} setup, the laser power used to excite the diamond was 0.239 mW and the collected fluorescence was \(0.309\ \mu \)W (the power meter was set at 635 nm).}
\label{fig: Figure10}
\end{figure}

As the NVs' fluorescence excitation and collection enhancing will greatly benefit NV-spin magnetometry applications \cite{Taylor2008, Pham2011}, the utilize of a TOF tip will certainly enhance its sensitivity. To give a hint on this enhancing effect, we resorted to CW-ODMR by detecting the fluorescence emission and varying the frequency of microwaves that induce spin transitions. We performed a series of experiments and compared the ODMR magnetic field detection sensitivity of two diamonds crystals bonded on TOF tip and flat fiber end (non-TOF). The results (Fig.\ref{fig: Figure9}) shown that under the same excitation laser power, the small $\sim$7.9-\(\mu \)m diamond on the 12.5-\( \mu \)m TOF tip (Fig.\ref{fig: Figure9} (a) ) have about 28.3 times the sensitivity of the $\sim$11.4-\(\mu \)m diamond on the flat fiber end. As a demonstration, we bonded a  $\sim$11.3-\(\mu \)m diamond on TOF tip by UV-curing glue and used it for measuring a DC magnetic field. We used only 0.238 mW of laser power to initialize the NV-centre ensemble, and -15 dBm of microwave power to induce the electronic spin resonance between the ms=0 spin state and the ms=$\pm$1 states of the NV centers. DC current was applied to a coil to vary the external magnetic field. The single shot ODMR spectra at various magnetic fields are shown in Fig.\ref{fig: Figure10}; the signals are obtained without utilizing any special techniques such as lock-in amplification or data averaging. 


\section{Conclusion}

To summarize, we have fabricated a quasi-adiabatic TOF tip that performs similarly to an objective in a classical optical path for both the excitation and collection of the fluorescence from NV centers in micrometer-sized diamond. Calculation shows that the TOF tip possesses an ultra-high NA; experiments demonstrate that this technique can greatly enhance the excitation and fluorescence collection efficiency. We used this enhancing to boost the magnetic sensitivity of micro-sized NV magnetometer and achieve a sensitivity of 180nT$\sqrt{Hz}$ for a $\sim$7.9-\(\mu \)m diamond crystal for DC magnetic field sensing. Furthermore, this improved excitation and collection enhancements reduce the size of the diamond sample that is required for precision sensing, potentially increases the spatial resolution (less than 5 \(\mu \)m)  and retains the flexibility of optical fiber-based NV sensors.\\ 

The authors thank Dr. Andrii Lazariev for critical reading of the manuscript. We gratefully acknowledge funding from the Max-Planck Society, Nieders\"achsisches Ministerium f\"ur Wissenschaft und Kultur and DFG Research Center Nanoscale Microscopy and Molecular Physiology of the Brain.

\clearpage 
\newpage
\pagebreak
\widetext
\begin{center}
\textbf{\large Additional Material}
\end{center}
\setcounter{figure}{0}
\makeatletter
\renewcommand{\thefigure}{Fig-add\arabic{figure}}
\renewcommand{\citenumfont}[1]{#1}

\begin{figure}[ht]
    \centering
    \subfloat{{\includegraphics[width=1.6in]{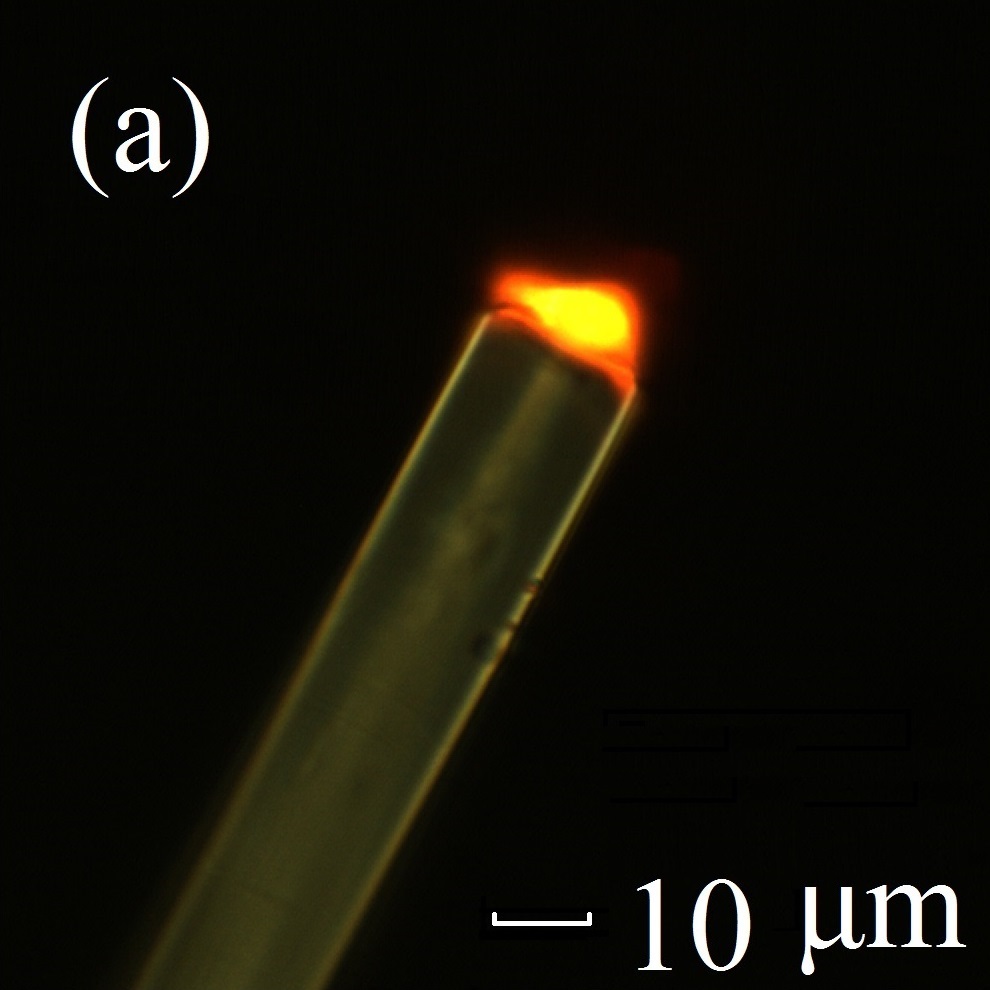} }}\hspace*{-1.5em}
    \qquad
    \subfloat{{\includegraphics[width=1.6in]{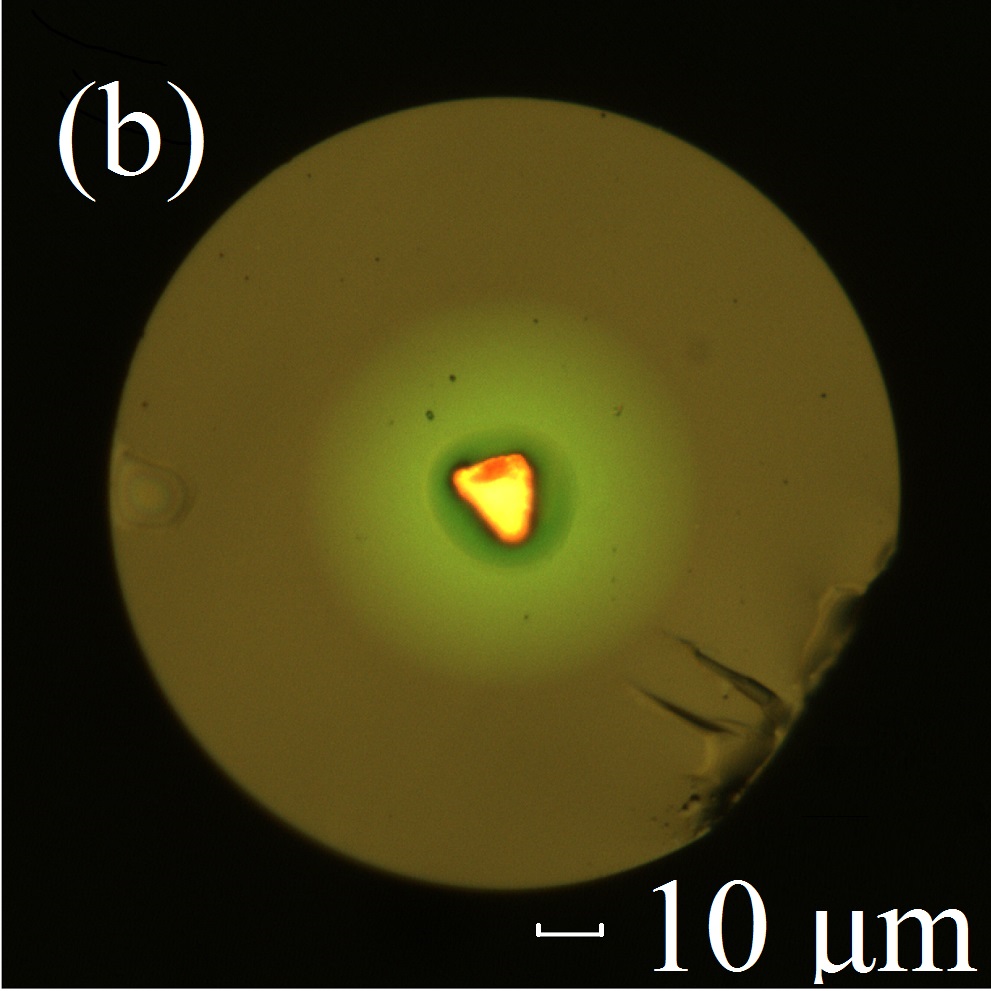} }}\hspace*{-1.5em}
 
    \subfloat{{\includegraphics[width=1.6in]{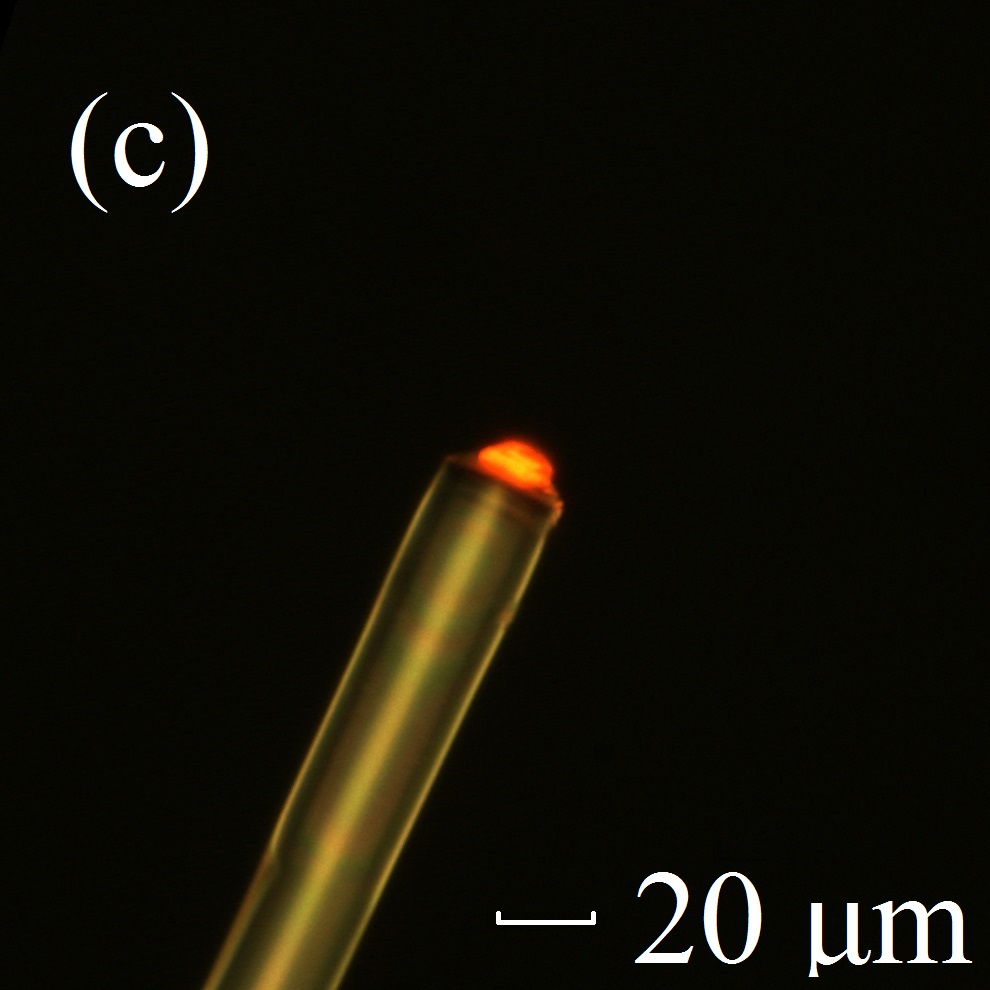} }}\hspace*{-1.5em}
    \qquad
    \subfloat{{\includegraphics[width=1.6in]{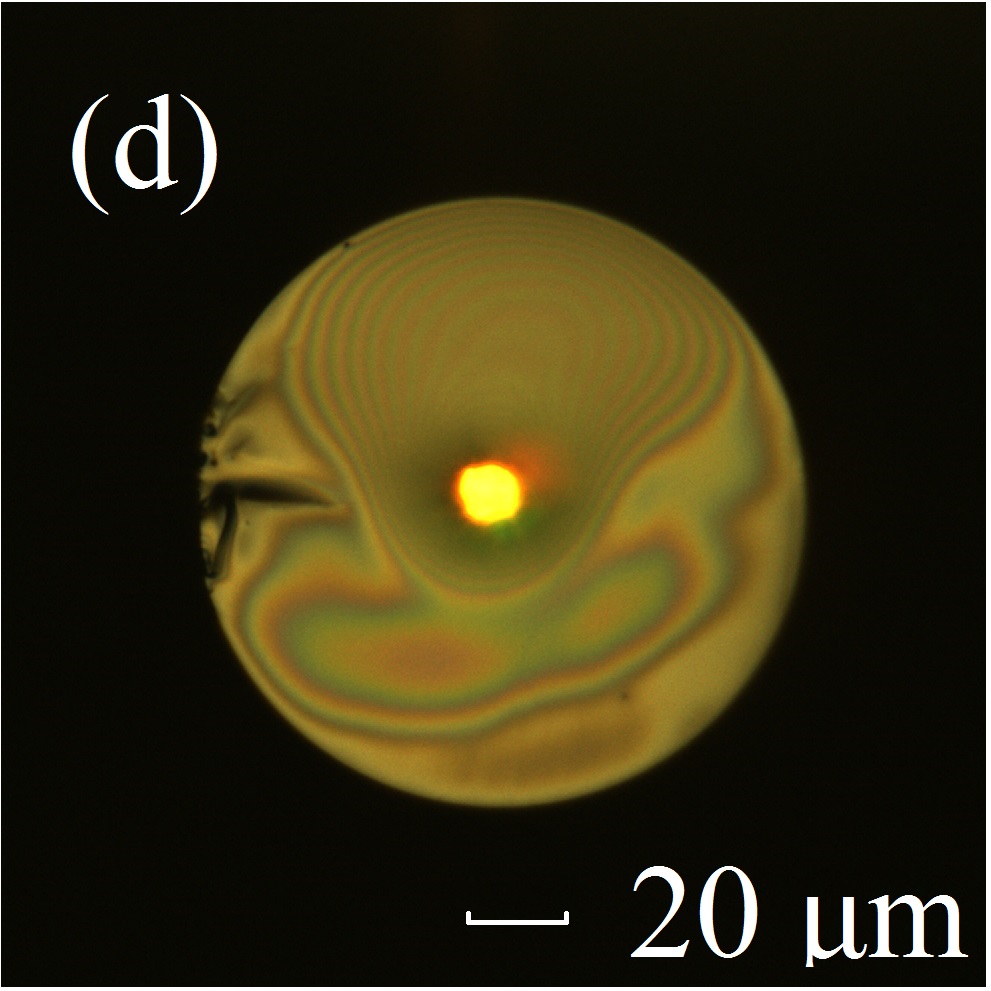} }}\hspace*{-1.5em}

    \caption{Microscope images of (a) the side view of the  $\sim$ 11.7-\(\mu\)m diamond on the TOF tip (the TOF tip has a diameter of  $\sim$16.8-\(\mu\)m), (b) the same diamond on the centre of the cleaved flat fiber end (the fiber diameter is  $\sim$125\(\ \mu \)m), (c) the side view of  the  $\sim$7.9-\(\mu\)m diamond bonded on the  $\sim$12.2\(\ \mu\)m TOF tip by UV glue, and (d) a  $\sim$11.4\(\ \mu \)m diamond bonded on the centre of the cleaved flat fiber end by UV glue.}
    \label{fig: Figures1}
\end{figure}

\section{Fluorescence collection simulation} 

\begin{figure}[ht]
\centering
\includegraphics[width=13.8cm]{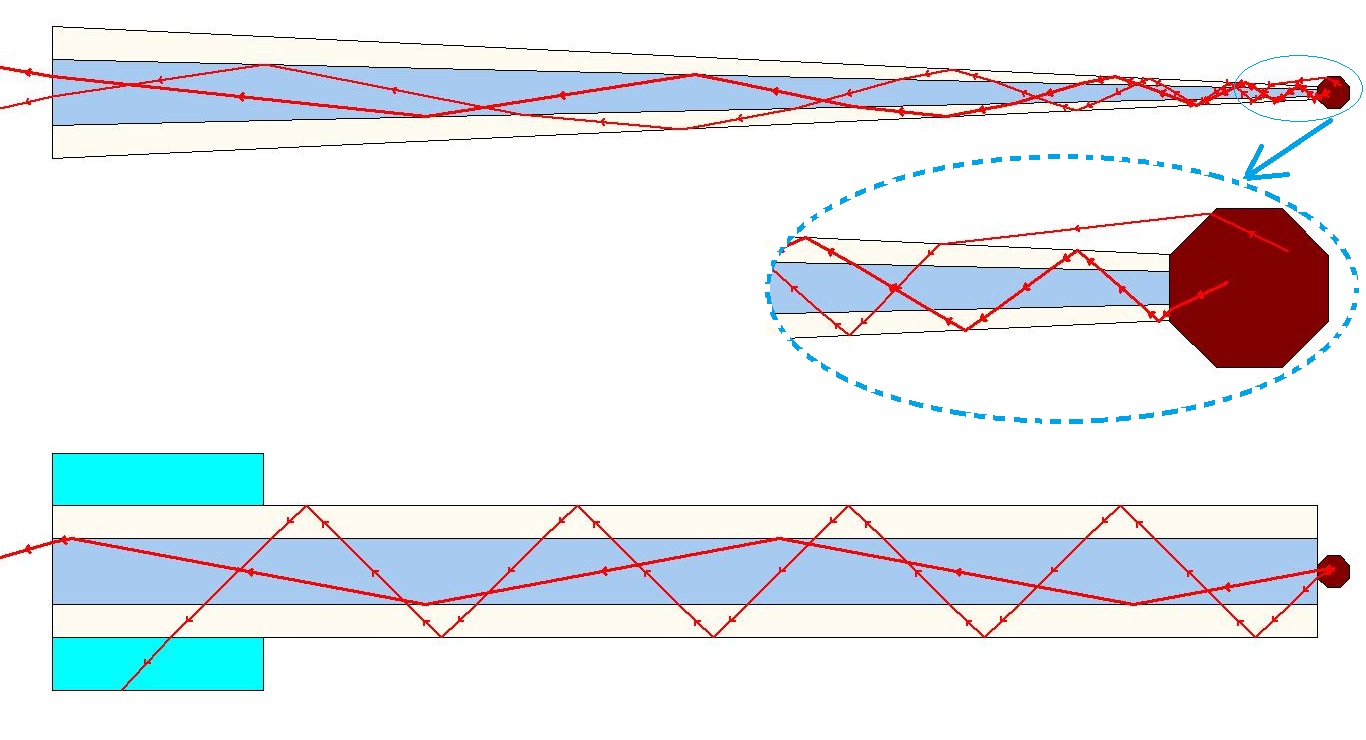}
\caption{Comparison of the simulation of the diamond NV center fluorescence collection of the(top)tapered optical fiber tip and (bottom)flat optical fiber end results.Simulation parameters: Fiber diameter is 125 \(\mu\)m,tapered optical fiber tip diameter is 12.5 \(\mu\)m,diamond diameter is 30 \(\mu\)m, length of the taper is 1.2 mm.}
\label{fig: Figures2}
\end{figure}

The simulation is carried out by free software Optgeo which can be downloded for free from http://jeanmarie.biansan.free.fr/optgeo.html.The results are shown in Fig.\ref{fig: Figures2}.a and b.

According to the comparison of the simulation of the diamond NV center fluorescence collection of the tapered optical fiber tip and flat optical fiber end results. The incidence angle of fluorescence for the tapered tip are both 26º, while for flat optical fiber tip, the incidence anges are 26º and 6.5º respectively. The refractive index of the diamond, fiber core, fiber cladding, polymer coating and air are 2.4, 1.496, 1.47, 1.56 and 1, respectively. The results shows that for the tapered optical fiber tip, the fluorescence have enter angle as large as 26º can be grasped by the fiber taper and guided into the optical fiber core through the tapered region, while that for the flat optical fiber end, only the fluorescence have incidence angle smaller than 6.5º can be guided in the fiber core, and the larger incidence angle light will be absorbed or  trapped out of the fiber by the polymer coating. 

\begin{figure}[ht]
\centering
      \includegraphics[width=5.2in]{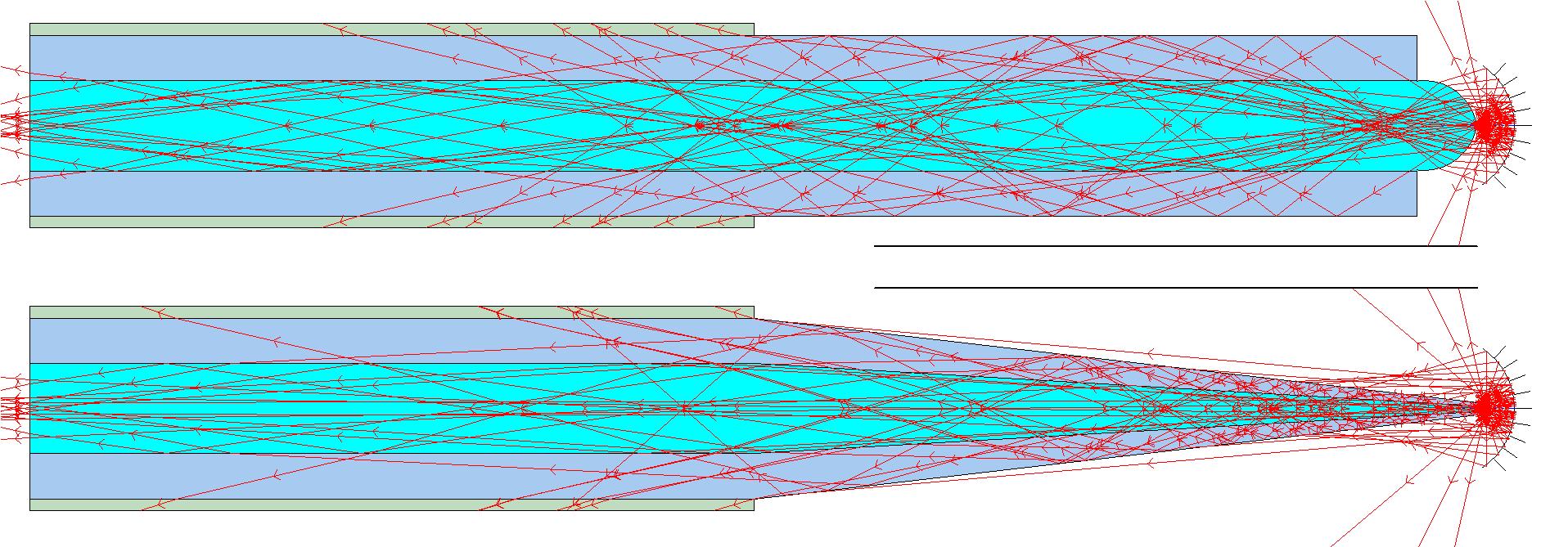}\\
      \caption{Comparison of the simulation of the diamond NV center fluorescence collection of the (a) sphered end combine with micro-concave mirror and (b) TOF tip combine with micro-concave mirror. Simulation parameters: Fiber diameter is 125 \(\mu\)m, tapered optical fiber tip diameter is 20 \(\mu\)m, micro-concave mirror curvature radius is 120 \(\mu\)m, length of the taper is 500 \(\mu\)m.}
\label{fig: Figures3}
\end{figure}

Fig.\ref{fig: Figures3}.b shows the simulation results of combining the micro concave technique (locate the diamond at the focal point of a micro-concave mirror facing the optical fiber tip) with the TOF tip, and Fig.\ref{fig: Figures3}.a shows the simulation results of using a sphered fiber end for comparison. It is clear that the micro-concave technique will enhance the collection efficiency for the TOF tip configuration.

\end{document}